# Long-lived state of a helium-like magnesium donor in silicon

**R.Kh. Zhukavin[1*], D.A. Postnov[1,2], P.A. Bushuikin[1], K.E. Kudryavtsev[1], K.A. Kovalevsky[1], V.V. Tsyplenkov[1], N.A. Bekin[1], A.N. Lodygin[3], L.M. Portsel[3], V.B. Shuman[3], Yu.A. Astrov[3], N.V. Abrosimov[4], V.N. Shastin[1]**

[1]Institute for Physics of Microstructures of the Russian Academy of Sciences, Academicheskaya Str., 7, Afonino, Nizhny Novgorod region, Kstovsky district, Kstovo region, 603087, Russia

[2]Lobachevsky State University of Nizhny Novgorod, Gagarin Ave. 23, Nizhny Novgorod 603022, Russia

[3]Ioffe Institute, Politekhnicheskaya St. 26, St. Petersburg 194021, Russia

[4]Institut für Kristallzüchtung (IKZ), Max-Born-Str. 2, 12489 Berlin, Germany

[*]zhur@ipmras.ru

## Abstract

The relaxation of ortho states of a helium-like Mg donor in silicon is investigated by measuring the modulation of background radiation transmission through impurity centers under pulsed photoexcitation. Long-lived states of the spin-triplet $1s(^3T_2)$ group with a lifetime of about 20 ms are observed. The temperature dependence indicates that the relaxation is governed by the Orbach mechanism with an activation energy $\Delta \sim 13$ meV, which is close to the exchange splitting energy of the excited $1s$ states of the Mg donor.

## Introduction

The use of donor and acceptor doping in semiconductors for the fabrication of various devices has required the development of numerous technological, experimental, and theoretical methods and approaches, which has significantly enriched solid-state physics. In recent years, considerable efforts have been directed toward the realization of quantum devices based on single impurity centers. At the present stage, this class of devices includes single-atom transistors [1], memory elements [2], and qubits [3]. In the latter case, as a rule, the use of electron spin states possessing sufficiently long relaxation times is assumed [4].

The most technologically advanced approach is based on group-V donors (most commonly phosphorus [5]) in monoisotopic silicon, which is associated with the well-developed technology of $^{28}$Si enrichment [6]. At low temperatures and low to moderate concentrations, and neglecting hyperfine interaction, the relaxation time for transitions between spin states of the $1s(A_1)$ orbital level of a hydrogen-like center is determined by Raman or Orbach processes [4, 7], whose efficiency depends on the spin–orbit interaction of the electron in the $1s(T_2)$ state and the corresponding energy gap. Deeper, helium-like donors have been studied to a much lesser extent, and the literature contains results only for singly ionized donors [8, 9], which is determined by the similarity of the experimental conditions required for these studies to those for hydrogen-like donors.

Spin-flip processes in two-electron systems can be observed only with the participation of ortho states, including ortho↔para and ortho↔ortho transitions. Since the ground state of the magnesium donor has total spin $S=0$, investigation of spin relaxation requires population of the ortho states in either equilibrium or nonequilibrium manner [10, 11].

Helium-like donors in silicon (magnesium, sulfur, selenium, and tellurium) possess two subsystems of energy levels that differ in total spin [12–14]. This feature appears promising from the viewpoint of creating devices that utilize spin-dependent states. In particular, one may envisage the possibility of realizing a single-atom spin filter based on spin-dependent relaxation. At present, however, data on the transition rates between spin-singlet and spin-triplet states of two-electron donors are not available.

Magnesium is an interstitial donor and, among doubly charged donors in silicon [15–17], has the lowest binding energy (107.5 meV, [18]) [Fig. 1(a)], with the exception of thermodonors [19]. Its relatively low atomic number allows one to expect a small magnitude of spin–orbit interaction, which should result in a suppression of spin-flip transitions. Magnesium is typically introduced into silicon by high-temperature diffusion [20–22], as a result of which both isolated magnesium centers and various molecular complexes can be formed; the latter may act as helium-like or hydrogen-like donors [21].

Earlier electron paramagnetic resonance experiments demonstrated that the $Mg^+$ ion has $T_d$ symmetry [8]. Piezospectroscopy studies made it possible to estimate the energy position of the excited $1s(E)$ state [21] of neutral $Mg^0$ to be $E\sim50$ meV; however, the spin component of the wave function was not discussed by the authors. Subsequently, absorption spectroscopy at temperatures around $T=100$ K and the observation of Fano resonances in the photocurrent spectrum allowed the energies of the spin-triplet states $1s(^3E)$ and $1s(^3T_2)$, as well as the spin-singlet state $1s(^1T_2)$, to be determined [22]. More recently, it was shown that resonant excitation of the para states $2p_0$ and $2p_\pm$ of neutral magnesium results in relatively fast relaxation times ($\sim10$–$20$ ps) observed in pump–probe measurements [23], with two-photon transitions playing an important role [24].

In the present work, the relaxation time of the lowest ortho state of the donor is investigated [Fig. 1 (a)], whose relaxation should be accompanied by a change in the spin component of the two-electron wave function. At present, neither a quantitative theoretical description of such relaxation nor a corresponding set of experimental data is available. Nevertheless, it is possible to rely on the extensive body of knowledge obtained for spin relaxation in hydrogen-like and singly charged helium-like donors to provide a qualitative description of the experimental results for the neutral two-electron magnesium donor.

As is well known [25–27], studies of spin relaxation times of hydrogen-like donors in silicon at low temperatures are performed in the presence of a magnetic field, which leads to Zeeman splitting of the energy levels. In the case of a two-electron system, spin–lattice relaxation can occur via transitions between spin-triplet and spin-singlet states even in the absence of a magnetic field. Below, we present the results of measurements of the relaxation time of a long-lived state of magnesium in silicon using modulation of continuous probe radiation under excitation of the donor by pulsed mid-infrared radiation.

**Experiment**

The samples studied were prepared by diffusion doping of single-crystalline silicon with magnesium [22]. The donor concentration was approximately $N_d \sim 3 \times 10^{15}$ cm$^{-3}$. The samples were cut in the form of a wedge (with an angle of about 1°), with an aperture of $5 \times 7$ mm$^2$ and a thickness of about 2 mm, and all faces were polished. Depending on the experiment, the samples were mounted either in a cryogenic insert placed in a helium Dewar vessel or in a closed-cycle optical cryostat with temperature control [Fig. 1(c)].

Measurements of the modulation of transmission of continuous Planck radiation at $T=300$ K were performed using a pulsed TEA CO$_2$ laser ($\lambda=10.6$ μm, $\hbar\omega=117$ meV) as the excitation source, with a peak power of up to $P=100$ kW, a pulse duration of 100 ns, and a repetition rate of 5 Hz. In addition, a Solar optical parametric generator (OPG) tunable in the range of 11–17 μm, with a peak power of up to $P=100$ kW, a pulse duration of 10 ns, and a repetition rate of 10 Hz, was employed.

To reduce the influence of the pump radiation on the Ge:Ga detector, a crystalline sapphire filter and the detector itself were positioned outside the propagation direction of the excitation beam [Fig. 1(*b*, *c*)]. When the optical cryostat was used, the detector temperature corresponded to that of the cryostat cold finger, whereas the sample temperature was varied using an additional heater. The thermal coupling was adjusted so that changes in the sample temperature within the measurement range did not affect the detector temperature or its sensitivity.

**Results and Discussion**

Absorption of a CO$_2$-laser radiation pulse during propagation through the sample leads to ionization of magnesium donors, followed by capture of free electrons and their relaxation, resulting in the formation of nonequilibrium populations of impurity states. At low and moderate excitation intensities ($I < 10$ kW/cm$^2$), a photomodulation signal is observed whose duration depends on temperature [Fig. 2]. The observed response corresponds to the appearance of induced absorption in the sample, that is, to a reduction in the conductivity of the Ge:Ga impurity detector used.

In contrast, high excitation intensities ($I > 10$ kW/cm$^2$) lead to a reversal of the detector signal sign, which corresponds to a dominance of radiation over absorption. Studies of spontaneous emission from Si:Mg under optical excitation have been reported previously [14], and the high-intensity excitation regime is therefore not considered in the present work.

Measurements of the temperature dependence of the photomodulation signal duration in Si:Mg reveal a strong variation in the temperature range of 4–15 K [Fig. 2], while no signal is observed at higher temperatures. At temperatures close to 15 K, the decay time decreases to about 2 μs, which corresponds to the bandwidth of the signal amplifier used. Similarly, the amplitude of the photomodulation signal decreases with increasing temperature.

In the temperature range $T=9.5$–14.5 K, two characteristic time components are observed in the signal [Fig. 2 (c)]. One of them (the "fast" component) exhibits a pronounced temperature dependence, whereas the other (the "slow" component) is nearly temperature independent and is likely associated with a lattice-related contribution whose duration does not depend on temperature. The photomodulation signal duration reaches up to 20 ms [Fig. 2 (b, c)] under liquid-helium cooling in an optical cryogenic insert placed in a Dewar vessel, when additional filters (sapphire and GaAs) are used to reduce background illumination of the sample.

Additional measurements were performed on undoped silicon samples as well as on phosphorus-doped silicon with a donor concentration of $N = 2 \times 10^{15}$ cm$^{-3}$. These samples did not exhibit temporal responses similar to those observed for Si:Mg in the same excitation intensity range.

The excitation spectrum of the induced-absorption signal [Fig. 3], measured using a tunable radiation source, exhibits a resonant feature near the transition to the $2p_0$ state, as well as an increase in the signal upon approaching the conduction band and a maximum in the photoionization region. No photomodulation signal is detected for excitation photon energies below that of the $1s(A_1) \rightarrow 2p_0$ donor transition. Since the energy of the $2p_0$ state lies close to the conduction-band edge, carrier emission from the $2p_0$ state into the continuum is likely to occur both via equilibrium phonons and through secondary absorption of pump photons. This indicates that the modulation of background radiation is formed during photoionization of neutral magnesium donors.

From pump–probe measurements of the relaxation times of spin-singlet states of magnesium, it is known that the corresponding times lie in the range of 10–20 ps [28]. It is therefore natural to assume that the rate of transitions between ortho states that are not accompanied by a change in the spin component of the wave function is of the same order of magnitude. Meanwhile, the characteristic temperatures at which the photomodulation signal is observed are rather low ($T$<15 K), and the relaxation times measured in the present experiment span a wide range from 2 μs to 20 ms in the temperature interval of 4–15 K. Such a broad range of relaxation times, together with their pronounced temperature dependence, indicates that the observed relaxation processes are governed by changes in the spin component of the wave function.

It should be noted that at low temperatures long relaxation times are observed for group-V donors in silicon under photoionization conditions. This behavior is associated with electron capture by neutral donors, resulting in the formation of negatively charged centers [29] with a binding energy on the order of one twentieth of the electron Bohr energy. However, for double donors, reports on the observation of negatively charged centers are absent in the literature. This is due to the impossibility of accommodating three electrons on a single $1s$ orbital, which would otherwise lead to the formation of states of the $1s2s2p$ type. As is well known from atomic physics, the binding energy of He$^-$ is an order of magnitude smaller than that of H$^-$ [30]. In the case of a helium-like donor in silicon, this yields a rough estimate of the binding energy on the order of 0.2 meV.

The absence of a photomodulation signal in magnesium-free samples allows the contribution of nonequilibrium phonons to be excluded. Thus, the characteristic time measured in the photomodulation experiment can be attributed to the relaxation time of the lowest ortho state $1s(^3T_2)$. The dependence of the photomodulation signal amplitude on the excitation intensity exhibits saturation already at relatively low intensities, which corresponds to the accumulation of electrons in the lowest ortho state.

The temperature dependence of the photomodulation signal duration $\tau$ in the range of 9–12 K (the "fast" component) is well described by an exponential function characteristic of the Orbach relaxation mechanism:

$$\frac{1}{\tau} \sim e^{-\frac{\Delta}{kT}} . \qquad (1)$$

This indicates the presence of an energy threshold for thermal activation of the relaxation process, whose value, according to the experimental results, is $\Delta$~13 meV. This value is close to the exchange-interaction-induced energy splitting between the spin-triplet $1s(^3T_2)$ and spin-singlet $1s(^1T_2)$ states [21]. The presence of spin–orbit interaction leads to coupling between the $1s(^3T_2)$ and $1s(^1T_2)$ states [12, 13].

Thus, relaxation of the electron in the lowest ortho state $1s(^3T_2)$ proceeds via the Orbach mechanism known for group-V donors, with the distinction that the spin–orbit interaction is present not only for the intermediate levels but also for the initial state [Fig. 1 (a)]. The existence of spin–orbit interaction for an electron in the $1s(^3T_2)$ state allows for the possibility of single-phonon relaxation to the ground state. However, within the range of experimentally observed relaxation times, no such direct process was detected.

As follows from the obtained data, the experimental points corresponding to measurements performed in liquid helium do not fall on the dependence obtained using the optical cryostat, which, in turn, exhibits saturation in the low-temperature region. This discrepancy is caused by limitations of the employed methodology: for sufficiently long relaxation times, background radiation begins to significantly affect the population of long-lived states. When a cryogenic insert placed in a Dewar vessel was used, the sample was exposed to substantially lower background illumination, which resulted in an increase in the observed relaxation time.

**Conclusion**

In this work, the relaxation of the lowest ortho state $1s(^3T_2)$ of a helium-like magnesium donor in silicon has been experimentally investigated using photomodulation of background radiation transmission under pulsed mid-infrared excitation. The absorption signal in the terahertz range is explained by the finite population of the spin-triplet $1s(^3T_2)$ state, whose long relaxation time is associated with the weak spin–orbit interaction.

The temperature dependence of the characteristic relaxation time in the range of 9–12 K is well described by the Orbach mechanism with an activation energy of $\Delta\sim 13$ meV, which is close to the exchange-induced energy splitting between the spin-triplet and spin-singlet excited 1s states of the magnesium donor. This indicates a crucial role of spin–orbit interaction coupling the $1s(^3T_2)$ and $1s(^1T_2)$ states in the relaxation process.

The absence of a photomodulation signal in magnesium-free samples allows the contribution of nonequilibrium phonons to be excluded and directly relates the observed relaxation times to the spin dynamics of the two-electron donor. The obtained results extend the understanding of spin relaxation mechanisms in helium-like donors in silicon and may be relevant for the development of devices exploiting spin-dependent states of single impurity centers.

The experimental studies were supported by the state assignment of IPM RAS (No. FFUF-2024-0019).

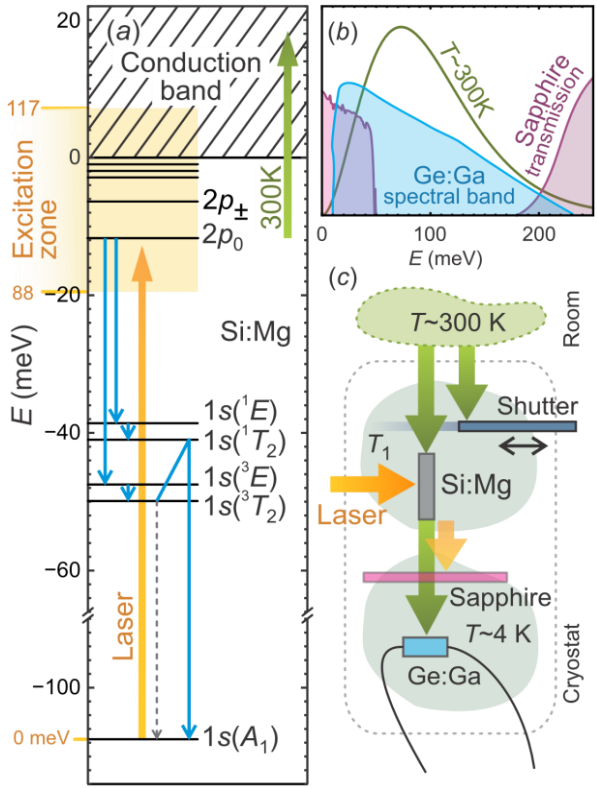

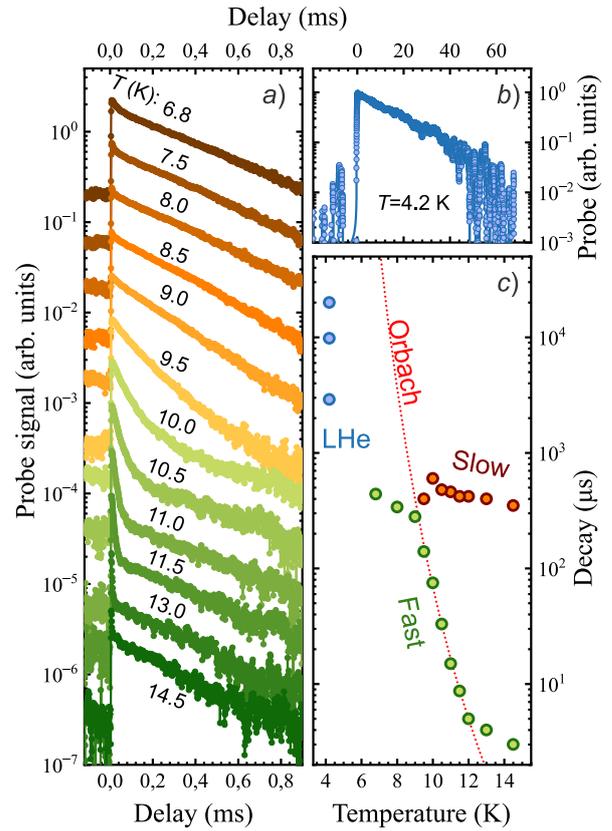

**PIC. 1** (*a*) Energy-level diagram of the doubly charged magnesium donor in silicon and the mechanism of excitation of nonequilibrium populations and relaxation of the electronic transition $1s(^3T_2) \rightarrow 1s(^1A_1)$. The dominant relaxation processes are indicated by solid arrows, while the direct relaxation channel is shown by a dashed arrow. (*b*) Sensitivity band of the Ge:Ga detector, transmission band of sapphire, and spectral density of room-temperature radiation. (*c*) Schematic of the experiment for measuring absorption modulation at different temperatures.

**PIC. 2** (*a*, *b*) Photomodulation signal of the background radiation during photoionization of magnesium donors in silicon at different temperatures. (*c*) Decay time of the background-radiation photomodulation signal as a function of temperature. The excitation pulse duration is $\tau_{CO2}$=100 нс ns, and the wavelength is $\lambda$=10.6 μm .

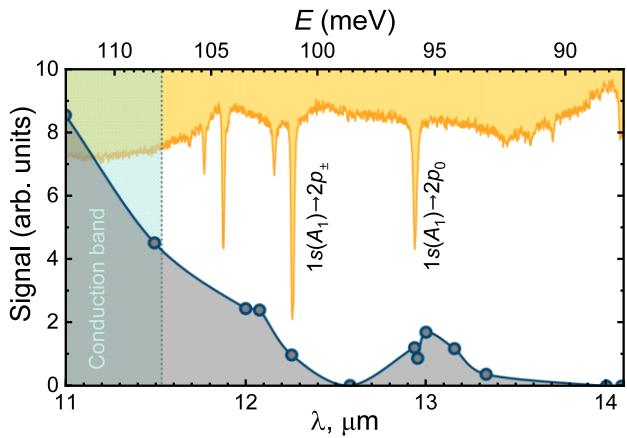

**PIC. 3** Absorption spectrum of Mg-doped Si and transmission-modulation signal versus excitation frequency of Mg donors at 4.2 K.